\documentclass[iop,numberedappendix]{emulateapj}

\usepackage{pslatex}
\usepackage{amsmath}
\usepackage{textcomp}

\newcommand{\myemail}{Gregory.A.Feiden.GR.@Dartmouth.edu}

\shorttitle{Mass-Radius Relation for Low-Mass MS Stars}
\shortauthors{Feiden \& Chaboyer}
\slugcomment{Accepted for Publication in the Astrophysical Journal}

\begin{document}

\title{Reevaluating the Mass-Radius Relation for Low-Mass, Main Sequence Stars}

\author{Gregory A. Feiden\altaffilmark{1} and Brian Chaboyer}
\affil{Department of Physics and Astronomy, Dartmouth College, 6127 Wilder Laboratory, Hanover, NH 03755}
\altaffiltext{1}{Neukom Graduate Fellow, \myemail}

\begin{abstract}
We examine the agreement between the observed and theoretical low-mass ($< 0.8\,M_{\odot}$) stellar main sequence mass-radius relationship by comparing detached eclipsing binary (DEB) data with a new, large grid of stellar evolution models. The new grid allows for a realistic variation in the age and metallicity of the DEB population, characteristic of the local galactic neighborhood. Overall, our models do a reasonable job of reproducing the observational data. A large majority of the models match the observed stellar radii to within 4\%, with a mean absolute error of 2.3\%. These results represent a factor of two improvement compared to previous examinations of the low-mass mass-radius relationship. The improved agreement between models and observations brings the radius deviations within the limits imposed by potential starspot-related uncertainties for 92\% of the stars in our DEB sample. 
\end{abstract}

\section{Introduction}
The disagreement between the theoretical and observational low-mass, main sequence mass-radius (henceforth MR) relationship has been recognized for nearly four decades \citep{Hoxie1970,Hoxie1973}. Although, only in the past two decades has the disagreement become overwhelmingly apparent with the reduction of observational uncertainties \citep[for an excellent review, see][]{Torres2010} and the development of sophisticated low-mass stellar models \citep{BCAH98}. 

The primary line of evidence stems from the study of detached double-lined eclipsing binaries (hereafter DEBs) with additional support garnered by direct measurements of stellar radii via interferometry \citep[e.g.,][]{Berger2006,vBraun2012}. These observations routinely suggest that stellar evolution models systematically under predict stellar radii by 5~--~15\% and over predict effective temperatures at the 3~--~5\% level. However, it is presently not clear whether the routinely quoted 5~--~15\% disagreement is representative of true radius discrepancies or whether there are other factors contributing to the derivation of such large radius errors. 

One such factor derives from the fact that previous studies focusing on the comparison between models and observations have generally applied a limited sample of isochrones to their data. Largely, these sets are comprised of 1~Gyr and 5~Gyr, solar metallicity isochrones. This is predominantly a consequence of the limited age and metallicity range of currently available low-mass stellar models. Age and metallicity effects are less important in the low-mass regime, but the stringent uncertainties quoted by observational efforts preclude the use of such a limited set of isochrones. For example, \citet{Burrows2011} discovered non-negligible radius variations in brown dwarfs and very-low-mass stars ($<0.1\,M_{\odot}$) when allowing for a more comprehensive set of metallicities. Isochrones with metallicities spanning a range characteristic of the local galactic neighborhood are therefore essential to accurately assess the validity of stellar evolution models.

Furthermore, one must also consider that the population of well-characterized DEBs has, until recently, consisted of eight systems. While unlikely, it is not unimaginable that those eight systems were more the exception than the rule in terms of their lack of consistency with stellar evolution models. Since publication of the \citet{Torres2010} review, the population of well-characterized, low-mass DEBs has more than doubled. The availability of this new data allows for a more accurate statistical characterization of the agreement (or lack of) between the MR relationship defined by models and observations.

When discrepancies are observed, they are typically attributed to the effects of a large scale magnetic field \citep[e.g.,][]{Ribas2006, Lopezm2007, Morales2008, Morales2009a, Torres2010, Kraus2011} as DEBs are often found in tight, short-period orbits with periods under three days. Tidal interactions and angular momentum conservation act to synchronize the orbital and rotational periods of the components, increasing the rotational velocity of each star in the process. The dynamo mechanism, thought to be responsible for generating and sustaining stellar magnetic fields, is amplified as a result of the rotational spin-up and enhances the efficiency of magnetic field generation within the star. Each component in the binary system is then more able to produce and maintain a strong, large-scale magnetic field than a comparable single field star. 

The effects of a large-scale magnetic field are thought to be two-fold: convective motions within the star are suppressed and the total surface coverage of starspots is increased. In both cases, a reduction in the total energy flux across a given surface within the star occurs, forcing the stellar radius to inflate in order to conserve flux \citep{Gough1966}. Recent attempts at modeling these effects have indicated that an enhanced magnetic field is a plausible explanation, although the primary physical mechanism affecting the structure of the star is still debated \citep{MM01,Chabrier2007,MM11}. 

Regardless of the precise physical mechanism, magnetic fields should betray their presence through the generation of magnetic activity in the stellar atmosphere. If magnetism is responsible for the observed inflated stellar radii, then correlations should be expected between individual stellar radius deviations and magnetic activity indicators (i.e., chromospheric H$\alpha$ and CaII H \& K emission, coronal x-ray emission, etc.). Tantalizing evidence of such correlations has been reported previously by \citet{Lopezm2007} and \citet{Morales2008}. 

However, recent evidence appears to stand in contrast with the current theory. Two systems, LSPM J1112+7626 \citep{Irwin2011} and Kepler-16 \citep{Doyle2011}, were discovered that have wide orbits with approximately forty-one day periods. Despite this, both appear to display discrepancies with stellar evolution models. In these systems, the component stars should be evolving individually with mutual tidal interactions playing a negligible role in the overall angular momentum evolution. The stars should be spinning down over time due to magnetic breaking processes \citep{Skumanich1972}, meaning the stars should not be as magnetically active compared with short period binary systems. The contrast is particularly evident for LSPM J1112+7626, where a rotation period of sixty-five days was detected via starspot modulation in the out-of-eclipse light curve. Gyrochronology suggests that the system has an age of approximately 9 Gyr \citep{Barnes2010} and implies further that the secondary is likely slowly rotating and should, therefore, not shown signs of strong magnetic activity or an inflated radius.

A third system also appears to defy the current hypothesis. KOI-126 \citep{Carter2011} is a hierarchical triple system with two low-mass, fully convective stars in orbit around a 1.35 $M_{\odot}$ primary. The two low-mass stars are orbiting each other with a period of 1.77 days. Therefore, they should show signs of inflated radii due to enhanced magnetic activity. However, it has been shown that the two low-mass, fully convective stars were in agreement with model predictions when considering their super-solar metallicity and the age of the higher mass primary \citep{Feiden2011}. This agreement was further confirmed by \citet{Spada2012}. 

Metallicity has been proposed previously as a solution to the observed MR discrepancies, but for the case of single field stars \citep{Berger2006}. This was contradicted shortly thereafter by \citet{Lopezm2007}, most notably for DEBs. Although, we must consider that the radius discrepancy-metallicity correlation is severely complicated by the fact that metallicities of M dwarfs are notoriously difficult to determine observationally.  

Finally, developments in light curve modeling of spotted stars has generated interesting results. The presence of large polar spots may alter the light curve analysis of DEBs by modifying the eclipse profile. These modifications lead to 2~--~4\% uncertainties in the derived stellar radii \citep{Morales2010, Windmiller2010, Kraus2011}. Thus far, only two DEBs (GU~Boo and CM~Dra) have been thoroughly tested for their sensitivity to spots. Systematic uncertainties may therefore dominate the error budget, casting a shadow of doubt on the observed radius discrepancies, which are often made apparent due to the minuscule random uncertainties. 

The uncertainties and developments outlined above have motivated us to reevaluate the current state of the low-mass MR relationship. In what is to follow, we use a large grid of theoretical stellar evolution isochrones in an effort to compare the low-mass models of the Dartmouth Stellar Evolution Program (DSEP) with DEB systems that have well constrained masses and radii. We then explore how potentially unaccounted for systematic uncertainties have the ability to create the appearance of discrepancies when neglected and mask real ones when considered. Section \ref{sec:data} will present the DEB sample followed by a description of the stellar models in Section \ref{sec:models}. The isochrone grid and fitting procedures will be explained in Section \ref{sec:proc}. Results will be presented in Section \ref{sec:result} followed by a discussion of the implications of our findings in Section \ref{sec:disc}. We conclude with a brief summary of the entire study in Section \ref{sec:summary}.

\section{Data}
\label{sec:data}
Our selection criteria mimic those of \citet[hereafter TAG10]{Torres2010} in so far as we require the random uncertainties in the mass and radius measurements to be less than 3\%. No data was disqualified due to perceived data quality issues or the original author's attempts to constrain possible systematic uncertainties. We also applied the criterion that the DEB system must include at least one component with a mass less than 0.8~M$_{\odot}$. This cut-off in mass is used to designate ``low-mass'' stars.

The mass cut-off was selected for two main reasons. First, the affects of age and metallicity on the structure of low-mass main sequence stars are suppressed compared to stars with masses of approximately 1.0~M$_{\odot}$ or above. Overall, this allows for less flexibility in fitting models to the observations, providing a more critical analysis of the stellar evolution models. Second, some of the largest discrepancies between observations and models are seen in the low-mass regime. This is likely a consequence of the former reason: true discrepancies become more apparent as the models become less sensitive to the input parameters.
	
Stars used in this study are listed in Table~\ref{tab:ebdata} along with their observationally determined properties and original references. Our final sample consisted of eighteen DEB systems for a total of thirty-six stars. Six of these systems are taken from TAG10 who reanalyzed the available data using a common set of reduction and parameter extraction routines in an effort to standardize the process. While the original references are cited, the parameters listed are those derived by TAG10 whose results were similar to the original values. 
	
A majority of the systems, ten in total, were published after the release of the TAG10 review. Six are drawn from the study performed by \citet{Kraus2011}, three are products of recent results from the \emph{Kepler Space Telescope} mission \citep[Bass et al. in prep.]{Carter2011, Doyle2011}, and the final post-TAG10 system was discovered by the M-Earth survey \citep{Irwin2011}. The remaining two systems from our sample were announced before TAG10 \citep{Lopezm2006,Lopezm2007b}, however, they were not included in the review for reasons related to either data availability or data quality. 
	
One final note: the Kepler systems KOI-126 \citep{Carter2011} and Kepler-16 \citep{Doyle2011} were not analyzed in a similar manner to the rest of the double-lined DEB population. They are DEB systems whose parameters were derived from Kepler photometry using a dynamical-photometric model \citep[see Supporting Online Material from][]{Carter2011}. However, they still satisfy the criteria for comparison with stellar evolution models and have been included for this reason.

\begin{deluxetable*}{l c c c c c}
\tablecolumns{6}
\tablecaption{DEB systems with at least one low-mass component with precise masses and radii.}
\tablehead{
\colhead{Star} & \colhead{$P_{\textrm{orb}}$} & \colhead{Mass} & \colhead{Radius} & \colhead{$P_{\textrm{rot}}$} & \colhead{Source\tablenotemark{$\dagger$}} \\
\colhead{Name} & \colhead{(day)} & \colhead{($M_{\odot}$)} & \colhead{($R_{\odot}$)} & \colhead{(day)} & \colhead{}
  }
\startdata
\object[UV Psc]{UV Psc A} &  0.86  & 0.9829 $\pm$ 0.0077   & 1.110  $\pm$ 0.023  & \nodata &  1 \\
UV Psc B      &        & 0.76440 $\pm$ 0.00450 & 0.8350 $\pm$ 0.0180 &  0.80   &    \\
\object[IM Vir]{IM Vir A} &  1.309 & 0.981  $\pm$ 0.012    & 1.061  $\pm$ 0.016  & \nodata &  2 \\
IM Vir B      &        & 0.6644 $\pm$ 0.0048   & 0.6810 $\pm$ 0.013  &  1.31   &    \\
KID 6131659 A & 17.528 & 0.924 $\pm$ 0.008     & 0.8807 $\pm$ 0.0017 & \nodata &  3 \\
KID 6131659 B &        & 0.683 $\pm$ 0.005     & 0.6392 $\pm$ 0.0013 & \nodata &    \\
\object[RX J0239.1-1028]{RX J0239.1-1028 A} &  2.072 & 0.7300 $\pm$ 0.0090   & 0.7410  $\pm$ 0.0040 & \nodata & 4 \\
RX J0239.1-1028 B  &        & 0.6930 $\pm$ 0.0060   & 0.7030  $\pm$ 0.0020 & \nodata &   \\
\object[Kepler-16]{Kepler-16 A} & 41.08  & 0.6897 $\pm$ 0.0034   & 0.6489 $\pm$ 0.0013 & \nodata &  5 \\
Kepler-16 B   &        & 0.20255 $\pm$ 0.0007  & 0.22623$\pm$ 0.0005 & \nodata &    \\
\object[GU Boo]{GU Boo A} &  0.49  & 0.61010 $\pm$ 0.00640 & 0.6270 $\pm$ 0.0160 &  0.49   &  6 \\
GU Boo B      &        & 0.59950 $\pm$ 0.00640 & 0.6240 $\pm$ 0.0160 &  0.54   &    \\
\object[YY Gem]{YY Gem A} &  0.81  & 0.59920 $\pm$ 0.00470 & 0.6194 $\pm$ 0.0057 &  0.87   &  7 \\
YY Gem B      &        & 0.59920 $\pm$ 0.00470 & 0.6194 $\pm$ 0.0057 &  0.82   &    \\
\object[MOTESS-GNAT 506664]{MG1-506664 A}  &  1.55  & 0.584 $\pm$ 0.002     & 0.560 $\pm$ 0.005   & \nodata &  8 \\
MG1-506664 B  &        & 0.544 $\pm$ 0.002     & 0.513 $\pm$ 0.009   & \nodata &    \\
\object[MOTESS-GNAT 116309]{MG1-116309 A}  &  0.827 & 0.567 $\pm$ 0.002     & 0.552  $\pm$ 0.004  & \nodata &  8 \\   
MG1-116309 B  &        & 0.532 $\pm$ 0.002     & 0.532  $\pm$ 0.004  & \nodata &    \\ 
\object[MOTESS-GNAT 1819499]{MG1-1819499 A} & 0.630  & 0.557 $\pm$ 0.001     & 0.569  $\pm$ 0.002  & \nodata &  8 \\   
MG1-1819499 B &        & 0.535 $\pm$ 0.001     & 0.500  $\pm$ 0.003  & \nodata &    \\ 
\object[NSVS 01031772]{NSVS 01031772 A} & 0.368 & 0.5428 $\pm$ 0.0027 & 0.5260  $\pm$ 0.0028 & \nodata &  9 \\
NSVS 01031772 B &       & 0.4982 $\pm$ 0.0025 & 0.5088  $\pm$ 0.0030 & \nodata &    \\
\object[MOTESS-GNAT 78457]{MG1-78457 A}   &  1.586 & 0.527 $\pm$ 0.002     & 0.505  $\pm$ 0.008  & \nodata &  8 \\   
MG1-78457 B   &        & 0.491 $\pm$ 0.001     & 0.471  $\pm$ 0.009  & \nodata &    \\
\object[MOTESS-GNAT 646680]{MG1-646680 A}  &  1.64  & 0.499 $\pm$ 0.002     & 0.457 $\pm$ 0.010   & \nodata &  8 \\
MG1-646680 B  &        & 0.443 $\pm$ 0.002     & 0.427 $\pm$ 0.008   & \nodata &    \\
\object[MOTESS-GNAT 2056316]{MG1-2056316 A} & 1.72   & 0.469 $\pm$ 0.002     & 0.441 $\pm$ 0.004   & \nodata &  8 \\
MG1-2056316 B &        & 0.382 $\pm$ 0.001     & 0.374 $\pm$ 0.004   & \nodata &    \\
\object[CU Cnc]{CU Cnc A} &  2.77  & 0.43490 $\pm$ 0.00120 & 0.4323 $\pm$ 0.0055 & \nodata & 10 \\
CU Cnc B      &        & 0.39922 $\pm$ 0.00089 & 0.3916 $\pm$ 0.0094 & \nodata &    \\
\object[LSPM J1112+7626]{LSPM J1112+7626 A} & 41.03 & 0.3946 $\pm$ 0.0023 & 0.3860 $\pm$ 0.0052 & \nodata & 11 \\
LSPM J1112+7626 B &       & 0.2745 $\pm$ 0.0012 & 0.2978 $\pm$ 0.0046 & \nodata &   \\
\object[KOI-126]{KOI-126 B} &  1.77  & 0.2413 $\pm$ 0.0030   & 0.2543 $\pm$ 0.0014 & \nodata & 12 \\
KOI-126 C     &        & 0.2127 $\pm$ 0.0026   & 0.2318 $\pm$ 0.0013 & \nodata &    \\
\object[CM Dra]{CM Dra A}   &  1.27  & 0.23102 $\pm$ 0.00089 & 0.2534 $\pm$ 0.0019 & \nodata & 13 \\
CM Dra B      &        & 0.21409 $\pm$ 0.00083 & 0.2398 $\pm$ 0.0018 & \nodata &
\enddata
\tablenotetext{$\dagger$}{(1) \citet{Popper1997}; (2) \citet{Morales2009b}; (3) G.~Bass et al. (in prep.);
 (4) \citet{Lopezm2007b}; (5) \citet{Doyle2011};
 (6) \citet{Lopezm2005}; (7) \citet{Torres2002}; (8) \citet{Kraus2011}; (9) \citet{Lopezm2006}; (10) \citet{Ribas2003};
 (11) \citet{Irwin2011}; (12) \citet{Carter2011}; (13) \citet{Morales2009a}; }
\label{tab:ebdata}
\end{deluxetable*}

\section{Models}
\label{sec:models}
Models utilized in this study were computed with the Dartmouth Stellar Evolution Program (DSEP)\footnote{http://stellar.dartmouth.edu/$\sim$models/}, a descendant of the Yale Rotating Evolution Code \citep{Guenther1992}. Physics included in the models have been described extensively in the literature \citep{Chaboyer2001,Bjork2006,Dotter2007,Dotter2008,Feiden2011}. Below, we will summarize the physics incorporated in the code that are pertinent for the present work.

Of particular importance for work on low-mass stars is the equation of state (EOS). In general, the EOS in DSEP is mass dependent. Stars with masses above $0.8 M_{\odot}$ are well represented by an ideal gas EOS with an appropriate Debye-H\"{u}ckel correction to account for ion charge shielding by free electrons \citep{Chaboyer1995}. Below $0.8 M_{\odot}$, non-ideal contributions to the EOS become non-negligible and must be considered. In this case, we use the FreeEOS\footnote{by Alan Irwin, http://freeeos.sourceforge.net} in the EOS4 configuration. FreeEOS allows for the treatment of an arbitrary heavy element abundance, providing DSEP with the flexibility to more reliably calculate models with super-solar abundances. 

Conditions in the outer, optically thin layers of low-mass stars precludes the use of grey atmosphere approximations and the use of radiative T($\tau$) relations \citep[and references therein]{CB00}. As such, it is critical to apply the results of non-grey model atmospheres to the definition of surface boundary conditions. For consistency, calculations across all mass regimes in this study define the surface boundary conditions using the PHOENIX AMES-COND model atmospheres \citep{Hauschildt1999a, Hauschildt1999b}. The atmospheres are attached at $T = T_{\textrm{eff}}$ by interpolating in tables generated by PHOENIX. Interpolation is performed in two variables, $\log g$ and $P_{\textrm{gas}}$, in order to define the temperature at the surface of the star. By attaching the atmospheres at $T = T_{\textrm{eff}}$, we eliminate the need to select an exact convective mixing-length value in the atmosphere calculations \citep{BCAH97}. However, this only applies to stars above $0.2 M_{\odot}$. Below that threshold, it becomes imperative to match the surface boundary conditions at a deeper optical depth in order to maintain an adiabatic atmosphere profile \citep{CB00}. We do not concern ourselves with this, for now, as none of the stars in our sample have a mass below the threshold.

Stars above about $0.35 M_{\odot}$ begin to develop radiative cores in our model set, with the size of the surface convection zone shrinking as mass increases. With this in mind, accurate radiative opacities are required. We use the OPAL high temperature opacities \citep{Iglesias1996} in conjunction with the low temperature opacities of \citet{Ferguson2005}. The Ferguson low temperature opacities are also utilized in the PHOENIX model atmospheres, providing consistency between the atmosphere calculations and the stellar envelope integration within DSEP.

As stars develop radiative cores it becomes increasingly important to include diffusion physics. Helium and heavy element diffusion are included \citep{Thoul1994}, unless the star is fully convective. Fully convective stars are assumed to be completely and homogeneously mixed throughout since the convective timescale dominates the diffusion timescale. When diffusion is calculated, a turbulent mixing term is included in the diffusion equations as prescribed by \citet{Richard2005}. The adopted reference temperature characterizing the efficiency of turbulent mixing is $\log(T_{\textrm{ref}}) = 6.0$ following the analysis of \citet{Korn2007}. 

Most of the stars in our sample may be considered rapid rotators. Therefore, one naturally wonders whether we should consider the effects of rotational deformation in our models. As a first approximation, we applied Chandrasekhar's analysis of slowly rotating polytropes to our DEB sample. \citet{Chandra1933} derived an expression for the stellar oblateness analytically. He defined slowly rotating to be when
\begin{equation}
\chi \equiv \frac{\omega^2}{2\pi \textrm{G} \rho_c} \ll 1
\end{equation}
where $\omega$ is the stellar angular velocity and $\rho_c$ is the central mass density. Assuming the stars in our sample are spin-orbit coupled, we find that all but two stars have $\chi < 10^{-4}$. The two exceptions (one DEB system) have $\chi\approx10^{-3}$; a consequence of their short orbital period of 0.368 days. Hence, all of the stars in our sample satisfy the slowly rotating approximation.

Based on the assumed polytropic index, n, Chandrasekhar derived that the deviation from sphericity (oblateness) could be approximated as
\begin{displaymath}
\mathcal{F} \equiv \frac{r_{eq} - r_{pole}}{r_{eq}} = \left\{
    \begin{array}{r c l}
        5.79\chi & & \textrm{for n = 1.5} \\
        9.82\chi & & \textrm{for n = 2.0} \\
        41.81\chi & & \textrm{for n = 3.0} \\
    \end{array}
    \right.
\end{displaymath}
with $\mathcal{F}$ being the oblateness, and $r_{eq}$ and $r_{pole}$ are the equatorial and polar radius, respectively. We approximated the polytropic index for each star using the interior density profile predicted by DSEP. The density profile for low-mass stars only slightly deviates from the polytrope prediction over the inner 98\% of the star (by radius). Specifically, we found that below 0.4~$M_{\odot}$, the models were best represented by an n~=~1.5 polytrope. Above 0.4~$M_{\odot}$ but below 0.65~$M_{\odot}$, an n~=~2.0 polytrope was appropriate. For all masses greater than 0.65~$M_{\odot}$, we defaulted to assuming an n = 3.0 polytropic index.

For all of the systems in question we found $\mathcal{F} < 0.001$, except for the one system that had the higher $\chi$ value. This system, NSVS~01031772, is slightly deviating from sphericity with $\mathcal{F} \approx 0.011$. We feel justified in neglecting the physics of rotation in our stellar models, with the caveat that attempting to probe model precisions below 1\% for NSVS~01031772 will likely require a more detailed treatment of rotational deformation.

Finally, we introduced the capability for DSEP to compute the characteristic convective overturn time. Our implementation is similar to the method of \citet{Kim1996}, who calculate the ``local'' convective overturn time at a distance above the base of the convection zone equal to one half of the mixing-length. This particular location was chosen based on the assumption that the tachocline (radiative-convective zone interface) is the source region of the stellar magnetic dynamo \citep{Parker1975}. However, using the stellar tachocline as our magnetic field source location for a fully convective star would be nonsensical. Therefore, in the case that the star is convective throughout, we utilize the results of \citet{Browning2008} as a first approximation to the magnetic field source location. Browning found that the magnetic field strength within a fully convective star was at a maximum at a depth of 85\% of the stellar radius. In accordance with this result, we compute the convective overturn time at one half the mixing-length above this location.

\begin{deluxetable}{l c c}
\tablecaption{Solar Calibration Parameters}
\tablehead{\colhead{Parameter} & \colhead{Adopted} & \colhead{Model}}
\startdata
Age (Gyr) & 4.57 & \nodata \\
$M_{\odot}$ (g) & 1.9891 $\times 10^{33}$ & \nodata \\
$R_{\odot}$ (cm) & 6.9598 $\times 10^{10}$ & $\log(R/R_{\odot}) = 8 \times 10^{-5}$ \\
$L_{\odot}$ (erg s$^{-1}$) & 3.8418 $\times 10^{33}$ & $\log(L/L_{\odot}) = 2 \times 10^{-4}$ \\
$R_{bcz}/R_{\odot}$ & 0.713$\pm$0.001 & 0.714 \\
$(Z/X)_{surf}$ & 0.0231 & 0.0230
\enddata
\label{tab:solar}
\end{deluxetable}

\section{Isochrone Fitting}
\label{sec:proc}

\subsection{Solar Calibration}
Carrying out a detailed comparison of theoretical models with observational data first requires the determination of a precise solar calibration configuration. This calibration is performed in order to determine the appropriate proto-solar helium mass fraction (Y), heavy element mass fraction (Z), and the convective mixing-length ($\alpha_{MLT}$). Using the heavy element abundance composition of \citet{GS98}, a $1.0 M_{\odot}$ model was evolved to the solar age \citep[4.57 Gyr,][]{Bahcall2005} at which point the model was required to reproduce the solar radius, solar luminosity, observed radius to the base of the convection zone, and the solar photospheric (Z/X). The final set of parameters required to match the solar properties were $Y_{init} = 0.27491$, $Z_{init} = 0.01884$, and $\alpha_{MLT} = 1.938$. Corresponding solar model parameters are listed in Table \ref{tab:solar}.

\subsection{Isochrone Grid}
\label{sec:isogrid}
Isochrones were computed for a wide range of age and metallicity values. The parameter space was defined to encompass a vast majority of stars typical of the local galactic neighborhood. Seven ages and seven scaled-solar metallicities were adopted for a total of forty-nine individual isochrones. The sets of values used in this study were: [Fe/H] = \{-1.0, -0.5, -0.3, -0.1, 0.0, +0.1, +0.2\} dex and age = \{0.3, 1.0, 3.0, 5.0, 6.0, 7.0, 8.0\} Gyr. For completeness, we also generated a set of isochrones that employed a smoothly varying convective mixing-length. However, for clarity, we defer the discussion of these models and relegate the information to Appendix~\ref{sec:vmixl}.

\subsection{Fitting Procedure}
\label{sec:iso_fit}
Judging agreement between observations and models was performed on a system-by-system basis, as opposed to fitting individual stars. Critical to the process was ensuring that both components of a given system were consistent with isochrones of a common age and metallicity. For each object, corresponding model radii were derived by linearly interpolating in each isochrone using the observationally determined mass. A linear interpolation scheme was sufficient since the mass resolution along the isochrones was small ($ \Delta M = 0.02 M_{\odot}$). 

Relative errors between the model radii and the observationally determined radius were then calculated. The relative error was defined as
\begin{equation}
\frac{\delta R}{R_{obs}} = \frac{R_{obs} - R_{model}}{R_{obs}}
\end{equation}
and will be presented as such throughout the rest of the paper. Formal agreement was determined by analyzing the number of standard deviations outside of the accepted range our model radii were located. Explicitly,
\begin{equation}
\#\, \sigma_{R} = \frac{R_{obs} - R_{model}}{\sigma_{R}}
\end{equation}
where $\sigma_{R}$ is the random uncertainty of the observational radius. 

Different levels of compatibility were assigned to each system based on the \emph{individual} agreement of each component in the system. For instance, if an isochrone matched both components at a common age within the $1\sigma$ limits set by the quoted random uncertainties, the system as a whole was designated as being ``fit'' by the models. Only within the $1\sigma$ level was a system considered to be accurately represented by the models.  The analysis was applied to two sets of data, which differ in the adopted radius uncertainties (formal quoted uncertainties or fixed 3\% uncertainties) that will be justified later. 

The process detailed above generated a list indicating the level of agreement between each isochrone and each DEB system. Narrowing the list to a single ``best fit'' isochrone involved minimizing the root mean square deviation (RMSD)
\begin{equation}
\textrm{RMSD} = \sqrt[]{\frac{1}{2}\displaystyle\sum_{i=1}^2 \left(\frac{\delta R_i}{\sigma_{R, i}}\right)^2}.
\end{equation}
where the sum is performed over the components of the DEB system. 

\subsection{Age \& Metallicity Priors}
\label{sec:prior}
DEBs that have been well studied have additional constraints that allow us to restrict the set of isochrones used in the fitting procedure. Specifically, our sample contains eight systems that have the added constraint of either an estimated age or metallicity, and in a couple cases, both. Before accepting the quoted age and metallicity priors, however, we performed a qualitative analysis to ensure that the estimates we adopted were reliable. A summary of our analysis is presented in Appendix~\ref{sec:appendix}. We judged four of the eight systems with quoted age or metallicity priors to have been determined reliably. Adopted priors are listed in Table~\ref{tab:prior}. 

\begin{deluxetable}{l c c c}[] 
\tablecaption{Age and [Fe/H] priors}
\tablehead{\colhead{DEB System} & \colhead{Age (Gyr)} & \colhead{[Fe/H]} & \colhead{Status} }
\startdata
UV Psc    & 7.9           & \nodata       & Rejected  \\
IM Vir    & \nodata       & -0.3$\pm$0.3  & Rejected  \\
Kepler-16 & \nodata       & $\le$ 0.0     & Accepted  \\
GU Boo    & $\le$ 1.0     & \nodata       & Rejected  \\
YY Gem    & 0.4 $\pm$ 0.1 & +0.1$\pm$0.2  & Accepted  \\
CU Cnc    & 0.3 $\pm$ 0.1 & \nodata       & Rejected  \\
KOI-126   & 4.0 $\pm$ 1.0 & +0.15$\pm$0.8 & Accepted  \\
CM Dra    & 4.0 $\pm$ 1.0 & $\le$ 0.0     & Accepted  
\enddata
\label{tab:prior}
\end{deluxetable}

\section{Results}
\label{sec:result}

\subsection{Direct Comparison}
\label{sec:result1}

\begin{deluxetable*}{l c r r r r c c r r r r} 
\tabletypesize{\scriptsize}
\tablecaption{Best fit isochrone with a solar calibrated mixing-length.}
\tablecolumns{12}
\tablehead{ 
 \colhead{} &  \multicolumn{5}{c}{Quoted} & & \multicolumn{5}{c}{Fixed 3\%} \\
 \cline{2-6}\cline{8-12} \\
 \colhead{Star Name} & \colhead{Age} & \colhead{[Fe/H]} & \colhead{$\delta R/R_{\textrm{obs}}$} & \colhead{\# $\sigma_{R}$} & \colhead{Fit} & &
  \colhead{Age} & \colhead{[Fe/H]} & \colhead{$\delta R/R_{\textrm{obs}}$} & \colhead{\# $\sigma_{R}$} & \colhead{Fit}
}
\startdata
UV Psc A          & 8.0 & -0.10 & -0.0285 & -1.375 &  No & & 8.0 & -0.10 & -0.0285 & -0.950 &  No \\
UV Psc B          &     &       &  0.1031 &  4.785 &     & &     &       &  0.1032 &  3.439 &     \\
IM Vir A          & 7.0 & -0.10 & -0.0037 & -0.248 &  No & & 7.0 & -0.10 & -0.0038 & -0.125 &  No \\
IM Vir B          &     &       &  0.0408 &  2.136 &     & &     &       &  0.0408 &  1.359 &     \\
KID 6131659 A     & 3.0 & -0.50 & -0.0014 & -0.750 & Yes & & 3.0 & -0.50 & -0.0014 & -0.047 & Yes \\
KID 6131659 B     &     &       &  0.0004 &  0.200 &     & &     &       &  0.0004 &  0.013 &     \\
RX J0239.1-1028 A & 8.0 & -0.10 &  0.0297 &  5.510 &  No & & 8.0 & -0.10 &  0.0297 &  0.991 & Yes \\
RX J0239.1-1028 B &     &       &  0.0285 & 10.007 &     & &     &       &  0.0285 &  0.949 &     \\
Kepler-16 A       & 1.0 & -0.10 &  0.0004 &  0.210 &  No & & 0.3 & -0.10 &  0.0099 &  0.330 & Yes \\
Kepler-16 B       &     &       &  0.0299 &  5.510 &     & &     &       &  0.0225 &  0.749 &     \\
GU Boo A          & 8.0 & -0.10 &  0.0292 &  1.144 &  No & & 8.0 & -0.10 &  0.0292 &  0.973 &  No \\
GU Boo B          &     &       &  0.0414 &  1.616 &     & &     &       &  0.0414 &  1.381 &     \\
YY Gem A          & 0.3 & -0.10 &  0.0813 &  8.834 &  No & & 0.3 & -0.10 &  0.0813 &  2.710 &  No \\
YY Gem B          &     &       &  0.0813 &  8.834 &     & &     &       &  0.0813 &  2.710 &     \\
MG1-506664 A      & 1.0 & -0.10 &  0.0035 &  1.951 &  No & & 1.0 & -0.10 &  0.0035 &  0.116 & Yes \\
MG1-506664 B      &     &       & -0.0031 & -1.612 &     & &     &       & -0.0031 & -0.105 &     \\
MG1-116309 A      & 7.0 & -0.50 & -0.0084 & -1.162 &  No & & 7.0 & -0.50 & -0.0084 & -0.281 & Yes \\
MG1-116309 B      &     &       &  0.0150 &  1.993 &     & &     &       &  0.0150 &  0.500 &     \\
MG1-1819499 A     & 8.0 & -0.50 &  0.0296 &  8.425 &  No & & 7.0 & -0.10 &  0.0404 &  1.348 &  No \\ 
MG1-1819499 B     &     &       & -0.0528 & -8.792 &     & &     &       & -0.0408 & -1.360 &     \\
NSVS 01031772 A   & 8.0 & -1.00 & -0.0021 & -0.396 &  No & & 8.0 & -0.50 & -0.0134 & -0.447 &  No \\
NSVS 01031772 B   &     &       &  0.0412 &  6.991 &     & &     &       &  0.0395 &  1.317 &     \\
MG1-78457 A       & 5.0 & +0.20 & -0.0008 & -0.049 & Yes & & 5.0 & +0.20 & -0.0008 & -0.026 & Yes \\
MG1-78457 B       &     &       & -0.0001 & -0.006 &     & &     &       & -0.0001 & -0.004 &     \\
MG1-646680 A      & 1.0 & +0.20 & -0.0226 & -1.725 &  No & & 1.0 & +0.20 & -0.0226 & -0.755 & Yes \\
MG1-646680 B      &     &       &  0.0184 &  1.309 &     & &     &       &  0.0184 &  0.613 &     \\
MG1-2056316 A     & 3.0 & +0.20 & -0.0074 & -1.632 &  No & & 3.0 & +0.20 & -0.0074 & -0.247 & Yes \\
MG1-2056316 B     &     &       &  0.0067 &  1.248 &     & &     &       &  0.0067 &  0.222 &     \\
CU Cnc A          & 8.0 & +0.20 &  0.0233 &  1.832 &  No & & 8.0 & +0.20 &  0.0233 &  0.777 & Yes \\
CU Cnc B          &     &       &  0.0020 &  0.084 &     & &     &       &  0.0020 &  0.067 &     \\
LSPM J1112+7626 A & 8.0 & +0.20 & -0.0005 & -0.035 &  No & & 8.0 & +0.20 & -0.0005 & -0.016 & Yes \\
LSPM J1112+7626 B &     &       &  0.0290 &  1.875 &     & &     &       &  0.0290 &  0.966 &     \\
KOI-126 B         & 1.0 & +0.10 & -0.0011 & -0.205 & Yes & & 3.0 & +0.20 & -0.0036 & -0.120 & Yes \\
KOI-126 C         &     &       & -0.0003 & -0.053 &     & &     &       & -0.0004 & -0.013 &     \\
CM Dra A          & 5.0 &  0.00 &  0.0316 &  4.221 &  No & & 5.0 &  0.00 &  0.0316 &  1.055 &  No \\
CM Dra B          &     &       &  0.0360 &  4.798 &     & &     &       &  0.0360 &  1.200 &  
\enddata
\label{tab:smixl}
\end{deluxetable*}

\begin{figure}
\plotone{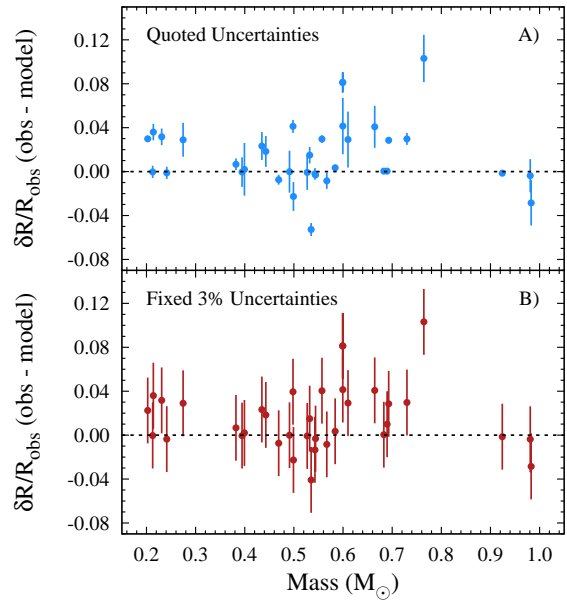}
\caption{Relative errors between stellar evolution models and observationally determined radii. Reliable age and metallicity priors were accounted for in the statistical analysis. \emph{(Top)} The adopted measurement uncertainties are the cited observational random uncertainties derived from the light curve fitting procedure. \emph{(Bottom)} A fixed 3\% uncertainty is adopted to represent possible systematic uncertainties (i.e., starspots).}
\label{fig:drad}
\end{figure}

Figure~\ref{fig:drad} demonstrates that, in general, our models reduce the observed radius discrepancies to below 4\% for 92\% of the stars in our sample. Only a few outliers are seen to be largely discrepant. Across the entire sample, we find a mean absolute error of 2.3\%. This broadly represents a factor of two reduction in the previously cited radius discrepancies. The age and metallicity of the best fit isochrone for each DEB system is listed in Table~\ref{tab:smixl} along with the level of agreement between the best fit isochrone and the individual stars comprising each DEB system. 

Contrasting with previous studies, we do not observe an overwhelming systematic trend of the models grossly under predicting stellar radii. Instead, systems discrepant by more than 5\% represent an exception to the broad agreement between the models and observations. While the agreement is admittedly not perfect, it is apparent that most observed discrepancies must now be judged according to the precision with which their radii were measured. As we will see in the next subsection, systematic uncertainties have the potential to blur our interpretation of the agreement between models and observations, simply due to the factor of two improvement described above.

Ignoring systematic uncertainties, for now, we are still confronted with the need to explain both the slightly discrepant as well as the largely discrepant radii. Discussion and comments pertaining to plausible explanations of the inflated radii will be addressed later in Section~\ref{sec:disc}.

\subsection{Fixed Uncertainties}
\label{sec:results2}
Recent work in modeling eclipse profiles of DEB systems has led to the realization that large, polar starspots can potentially affect observationally derived radii and subsequently dominate over the quoted random uncertainties. \citet[GU Boo;][]{Windmiller2010} and \citet[CM Dra;][]{Morales2010} have found that systematic uncertainties may be present in DEB radii on the order of 2~--~4\%. This is to be compared with quoted random uncertainties that are typically on the order of 1\% or less. 

Similarly, \citet{Kraus2011} included an estimate of systematic uncertainties for their radius measurements and found typical uncertainties on the order of 2~--~3\%. With the radius uncertainties potentially dominated by often unquoted systematics, we were curious to see what effect such uncertainties would have on comparisons with theoretical models. Quite obviously, observed discrepancies would be reduced by the level of systematic uncertainty. However, it has not been clear whether those systematics have the ability to completely mask the observed radius residuals, particularly when combined with isochrones that cover a wide age-metallicity parameter space. Since the radius uncertainty cut-off for our DEB sample was set at 3\%, we elected to adopt a fixed 3\% radius uncertainty in our analysis to mimic potential systematics.

Demonstrated in Figure~\ref{fig:drad} B is the effect of including a fixed 3\% radius uncertainty. The overall distribution of points appears to be similar to the previous case, except that the larger uncertainty enables more systems (eleven in total) to be considered fit by our analysis (see Table~\ref{tab:smixl}). This result is clearly expected when introducing larger error bars, as mentioned above. 

What is important to realize is that we are now presented with a scenario where the systematic uncertainties mask our ability to draw firm conclusions about whether the observed radius discrepancies are inherently real or merely a consequence of neglected uncertainties. Such ambiguity was previously not present since the observed radius residuals were substantially larger than any potential systematic uncertainties. 

\subsection{Peculiar Systems}
Before continuing with a further interpretation of our results, we wanted to briefly comment on several individual DEB systems. These systems stood out in our mind as worth remarking upon separate from the ensemble.

Our models never fit the UV Psc system to a coeval age. While we were able to fit UV Psc~A, the secondary component was always found to have an 8~--~10\% larger radius than would be expected from stellar evolution theory, consistent with \citet{Popper1997}. Both stars may realistically be discrepant with the models, however, if the age is truly younger than 8~Gyr. Constraining the actual magnitude of the discrepancies is difficult without observational age and metallicity estimates. Fittingly, the system is known to be very active based on spectroscopic analysis of H$\alpha$ cores, multi-epoch photometric monitoring \citep{UVPsc2005}, and the derived x-ray luminosity (this work, Section~\ref{sec:xray}). Further observations of UV~Psc to constrain its metallicity and to provide data for a more rigorous starspot analysis would be a worthwhile endeavor. 

MG1-189499, characterized by \citet{Kraus2011}, was also never found to be in total agreement. The primary and secondary were found to deviate by 3.0\% and -5.3\%, respectively. This appeared to be a cause for concern, but Kraus et al. were one of the few groups to provide an estimate of potential systematic uncertainties. Uncertainties in the primary star's radius were elevated to 4.6\% with a comparatively large uncertainty of 3.4\% in the secondary's radius. Applying these uncertainties and rerunning the isochrone fitting procedure allows the predictions of the models to nearly fall inline with the observations. An age of 8 Gyr was again derived but with a new metallicity of [Fe/H]~=~-0.3. This isochrone yielded relative radius errors of 4.8\% and -2.9\% for the primary and secondary, respectively. Hence, the secondary is now considered fit by the models and the primary is only 0.2\% outside the bounds of uncertainty.

The last two systems we would like to discuss are two that have long posed problems for modelers: YY~Gem and CM~Dra. YY~Gem \citep{Torres2002} is an equal mass, equal radius DEB system that is effectively represented by a single point at (0.6, 0.08) in Figure~\ref{fig:drad}. Beginning with the low-mass models of \citet{Hoxie1970}, YY Gem has never been adequately reproduced by stellar models.

The difficulty with YY Gem seems to be in the estimated age of 400~Myr, determined by its association with the Castor AB quadruple system. An older age is logically preferred by the models considering the notably inflated radius. Interestingly, this would have led previous studies to underestimate the observed radius discrepancy. Assuming either a 1 or 5 Gyr isochrone naturally results in models with larger radii than those computed with an age of 400~Myr. Further study of this system, particularly its distance (Section~\ref{sec:xray}) and association with Castor AB (see Appendix~\ref{sec:appendix}), will be beneficial toward fully understanding the nature of its inflated radii.

Finally, CM Draconis is another constant thorn in theoreticians sides. Here, the primary and secondary deviate by 3.2\% and 3.6\%, respectively. CM Dra is one of the most studied DEB systems and has very well constrained physical properties measured from data spanning multiple decades \citep{Morales2009a}. Our models indicate that a solar composition is preferred\footnote{A super-solar metallicity is actually favored, as was also noted by \citet{Spada2012}, but the application of the metallicity prior in our analysis restricted the models to only solar or subsolar compositions.}. However, the composition of CM Dra, while not known positively, is very likely subsolar \citep{Viti1997,Viti2002,Morales2009a,Kuznetsov2012}, meaning our models are probably more discrepant than indicated by this study. \citet{Morales2010} found that large polar spots produce systematic uncertainties in the radius measurements of around 3\%, potentially bringing CM Dra more in line or more out of line with model predictions. For the time-being, this system will continue to test our knowledge of stellar evolution.

\begin{figure*}
\plotone{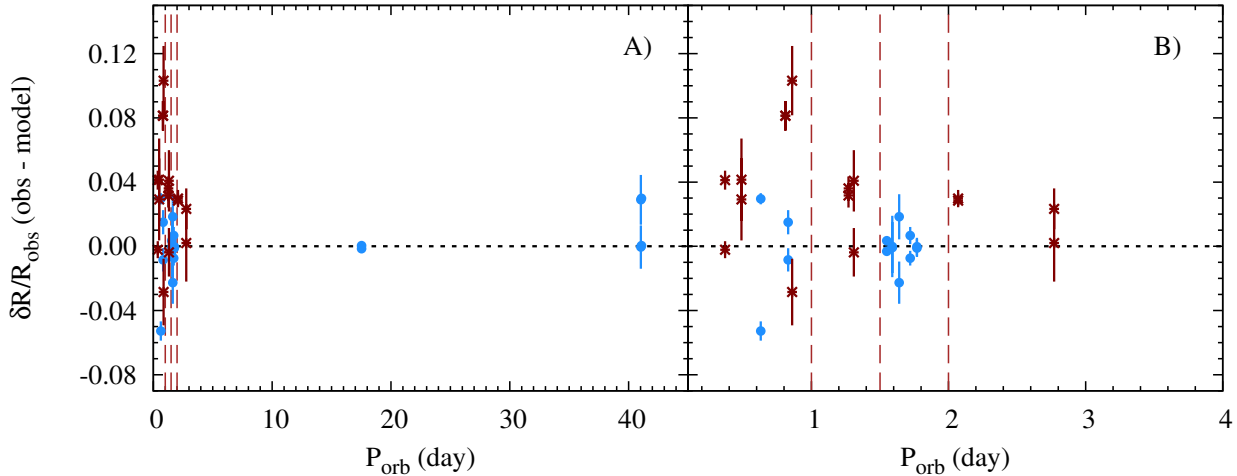}
\caption{Observed correlation between radius deviations and DEB orbital periods. Maroon asterisks are stars with known x-ray flux measurements. Dashed vertical lines represent the various period cuts used in our statistical analysis to divide the sample into two subsamples. \emph{(Left)} The full range of periods present in the current sample. \emph{(Right)} Highlighting the short period regime as it encompasses a majority of the DEB systems.}
\label{fig:porb}
\end{figure*}

\section{Discussion}
\label{sec:disc}
Data presented in this study hint at two competing explanations for the occurrence of the differing model and observational MR relations. It is entirely plausible that stellar evolution models are not incorporating key physical processes that can account for the observed discrepancies. This view is not new and has always accompanied discussions of low-mass models \citep{Hoxie1970,Hoxie1973,CB97,BCAH97,BCAH98}. Typically cited is our incomplete knowledge dealing with the complex array of molecules present in M-dwarf atmospheres as well as the lack of structural changes induced by a large-scale magnetic field. Included in the latter are both the effects on convective energy transport and the emergence of spots on the stellar photosphere.

The other scenario is one in which the neglect of systematic uncertainties is driving the apparent discrepancies. We have shown that by allowing for realistic variation in age and metallicity of the models, the radius residuals are of the same magnitude as the potential systematic uncertainties. Note, this is without any modification to the solar calibrated mixing-length or the inclusion of any non-standard physics. Previous studies have indicated that systematics may help alleviate the size of the radius residuals, but have required additional modifications to the models in order to fully reconcile models with observations. It is now clear that we must work to constrain and minimize systematic uncertainties in observations of DEBs to allow for them to provide an accurate test of stellar evolution models. 

\subsection{Radius Deviations}
The MAE between our models and the observations was 2.3\%, a factor of two improvement over the canonical \emph{minimum} of 5\%.  We found deviations of no more than 4\%, with the exception of a few stars, instead of ubiquitous 5~--~15\% errors, as is often quoted. Despite improving the situation, panel~A of Figure~\ref{fig:drad} illustrates that the models are still unable to fully reproduce the observed stellar radii. 

Accepting the factor of two improvement presented in Section~\ref{sec:result1}, the paradigm of broad disagreement between models and observations is shifted to one where agreement is broad, and large discrepancies are an exception. With the radius deviations typically less than 4\%, an evaluation of the systematic errors becomes imperative. Formerly, systematic uncertainties of about 4\% were incapable of relieving radius deviations greater than 5\%. Stellar evolution models still appeared to disagree with DEB observations even after the inclusion of systematic errors. 

The reason for the factor of two improvement is twofold. First, we have calculated models with a finer grid of metallicities. DSEP utilizes an EOS that enables models to be more reliably calculated for super-solar metallicities, allowing for a greater range of stellar compositions to be considered. Low-mass stellar models with super-solar metallicity have previously been unavailable for comparison with low-mass DEB data. Thus, the ability to extend our model set to super-solar metallicities allows for more flexibility in attempting to match the observed properties of DEB components.

Second, a far larger number of low-mass DEBs with precisely measured radii were available to us as opposed to previous studies. Before the publication of TAG10, there were only eight systems that met the criteria necessary to accurately constrain stellar evolution models. Following the TAG10 review, the total number of systems that met the necessary criteria more than doubled with the addition of ten newly characterized DEBs. These additional systems appear to be more in line with the results of standard stellar evolution theory. However, well-known discrepant systems remain noticeably discrepant and still require further explanation.

We must now ask, ``what belies the current discrepancies between models and observations?'' Figure~\ref{fig:drad} favors the hypothesis that non-standard physics are absent from current stellar evolution models. Our larger data set allows us to notice that stars of similar masses from different DEB systems appear to be discrepant at varying levels, an effect a single set of standard models cannot correct. However, this is contingent upon the accuracy our age and metallicity predictions. Until we have better empirical age and metallicity estimates for the various systems, it is too difficult to ascertain the true level of discrepancy for any individual star.

The efficiency of convective energy transport is of greatest interest. It is possible that convection is naturally inefficient. Although, we gather from Figure~\ref{fig:drad} that suppression of convective energy transport must be tied to a stellar property that is largely independent of mass. Simple parametrization of the suppression of convection is too uniform over a given mass regime to fully account for the observed differences in stellar radii for stars with similar masses (see also Appendix~\ref{sec:vmixl}). Any effort, either theoretical or observational, to constrain the physics of convection in low-mass stars will lend crucial insight.

The most favored option, is that convection is not intrinsically inefficient, but that a large-scale magnetic field acts to suppress convective motions \citep{MM01,Chabrier2007,MM11}. Stellar evolution models self-consistently incorporating the effects of large-scale magnetic fields will help on this front. Observations of cool-star magnetic field strengths and topologies will then provide a means of judging the validity of any new models. 

One final hypothesis is that starspots may affect the structure of stars and generate the inflated radii we observe. \citet{Chabrier2007} investigated such a possibility by artificially reducing the total stellar bolometric luminosity in an effort to mimic the effects of spots. They found that radius discrepancies were relieved with their parametrization. However, it is still not apparent whether spots reduce the total bolometric flux or if they locally shift flux to longer wavelengths \citep{Jackson2009}, preserving the total luminosity. Ultimately, if starspots are required in stellar evolution models, their inclusion is necessitated in the analysis of DEB light curves.

Quantifying the effect starspots may have on observed DEB light curves is extremely difficult. Obtaining accurate knowledge of the total surface coverage of spots, the total number of spots, their individual sizes, temperature contrasts, and their overall distribution on the stellar surface is nearly an impossible task given only a light curve. From a theoretical perspective, finding a proper parametrization to mimic the effect of spots on a three dimensional volume within the framework of a one dimensional model provides its own complications. Currently, the only feasible method to include spots, is to include their potential effects on the radius measurement uncertainties.

Unfortunately, while the inclusion of fixed 3\% radius uncertainties in our analysis was able to alleviate many of the noted radius discrepancies, it also created a situation where the measurement uncertainty was on the order of the typical radius deviation. We are presented with a case where the observations are no longer effective at testing the models and the manifestation of most radius discrepancies can be attributed to under estimated error bars. At this point, we require observations that have been rigorously vetted for potential systematic uncertainties and are able to still provide mass and radius measurements to better than 2\%.  

\subsection{Radius-Rotation-Activity Correlations}
Direct measurements of low-mass magnetic field strengths are rare, especially among fast rotating stars \citep{Reiners2012}. Without a direct measure of the magnetic field strength, we are forced to rely on indirect measures to probe correlations between stellar magnetism and the appearance of inflated stellar radii. Ideally, these indirect measures are intimately connected with the dynamo mechanism, thought to generate and maintain stellar magnetic fields, or are the product of magnetic processes in the stellar atmosphere. The preferred indirect measures are typically stellar rotation or the observation of magnetically driven emission (H$\alpha$, CaII H and K, X-ray).

\subsubsection{Rotation}
\label{sec:rot}
Typically, low-mass DEBs are found in tight, short period orbits ($<$~3d; see Figure~\ref{fig:porb}). Tidal interactions spin-up the individual components and allow them to remain rapidly rotating throughout their life cycle. Stellar dynamo theory dictates that the large-scale magnetic field strength is tied to the rotational properties of a star \citep[i.e., a rapidly rotating star should be more magnetically active than a comparable star that is slowly rotating;][]{Parker1979,Reiners2012b}, providing a natural starting point for our investigation. 

We performed two independent statistical tests on the distribution of radius residuals as a function of the orbital period ($P_{\textrm{orb}}$). Our primary objective was to determine whether rapidly rotating systems produce, on average, larger radius deviations than systems perceived to be slow rotators. Rotational periods were assumed to be synchronous with the orbital period unless a separate value for the rotational period was cited in the literature. The statistical tests performed were a Kirmogorov-Smirnov (K-S) test and another whereby we tested the probability of obtaining a given distribution of residuals via a Monte Carlo method. Both tests were performed on the observed difference in mean absolute error (MAE)\footnote{We selected the MAE over the RMSD as a measure of the mean radius deviation of a given ensemble in order to reduce the weight of any individual outlier in the final mean.} between two data bins. The data bins were divided at preselected values of $P_{\textrm{orb}}$, identified visually as vertical dashed lines in Figure~\ref{fig:porb}.

Comparing the radius deviations with the rotational periods, we find no evidence of any dominant correlation. Figure~\ref{fig:porb} displays the residual data as a function of the orbital period, with frame A showing the full range of observed periods and frame B highlighting the ``short period'' regime. The correlation of radius deviations with orbital period has been studied previously \citep[e.g.,][]{Kraus2011} where a significant difference between the two bins was observed around $P_{\textrm{orb}}$ = 1.5 days. We performed the statistical tests using three values for the orbital period (1.0d, 1.5d, and 2.0d) that defined the two period bins.

\begin{figure}
\plotone{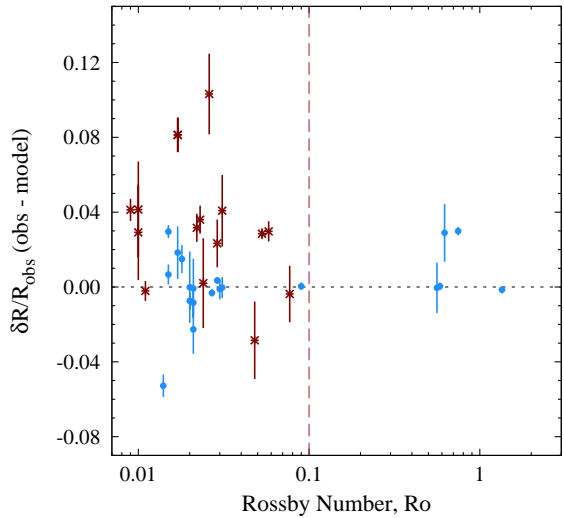}
\caption{The theoretical Rossby number, Ro = $P_{\textrm{rot}}/\tau_{\textrm{conv}}$, versus the relative radius error. Ro is tied directly to the theoretical stellar dynamo mechanism and is empirically related to the ratio of a star's x-ray to bolometric luminosity (a magnetic activity indicator). Asterisks in maroon are stars with known x-ray flux measurements.}
\label{fig:rossby}
\end{figure}

\begin{figure*}[!ht]
\epsscale{0.95}
\plotone{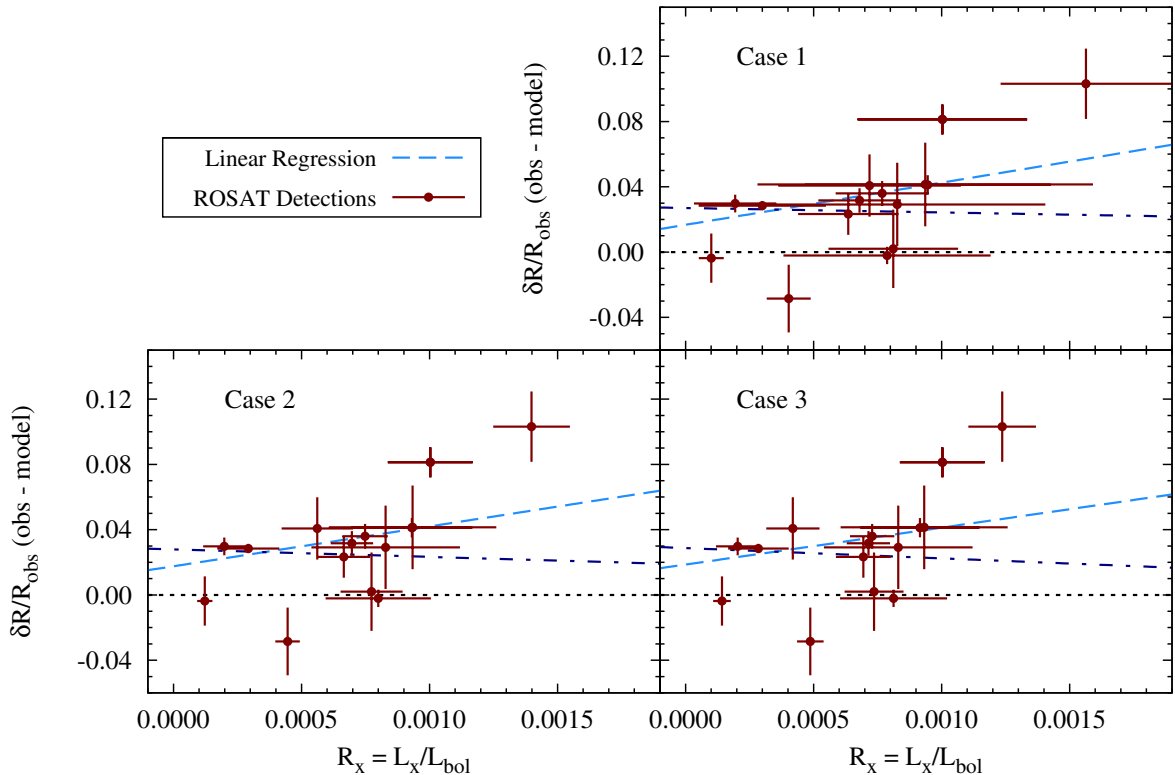}
\caption{Relative error between observational and model radii for stars with detected x-ray emission. X-ray data is drawn from the \emph{ROSAT} All-Sky Survey and combined with the radius residuals derived from this study. Data are shown as maroon filled circles. Illustrated are two least-square regressions performed on the data. The light-blue, dashed line demonstrates a non-negligible slope of $\sim25\pm17$ across all of the data, while the indigo, dash-dotted line excludes the two most discrepant points, UV Psc B and YY Gem.}
\label{fig:xray}
\end{figure*}

We confirm the results of \citet{Kraus2011} and find a $3.1\sigma$ difference in the distribution of radius deviations around 1.5d. Systems with $P_{\textrm{orb}} < 1.5$d had a MAE of 3.4\% while the longer period systems had a MAE of 1.0\%. While this is tantalizing, we can not necessarily attribute any physical significance to this particular division. We should expect this difference to be present for any two subsamples. However, we fail to find any evidence for a statistically meaningful difference when we divide the subsamples at 1.0d and 2.0d. Therefore, the significant difference noted at 1.5d is likely a spurious statistical result\footnote{\citet{Kraus2011} posit the difference may actually be a by-product of the DEB light curve analysis methods. Providing a further examination is outside the scope of this study.}. Inclusion of more long-period DEBs will be instrumental in providing a robust conclusion. Until those systems are discovered, there does not appear to be a physically meaningful explanation for why the divide should be made at 1.5d, but then also not hold for a division at 1.0d. 

The rotational (or orbital) period is not necessarily the most appropriate proxy for the magnetic field strength or potential magnetic activity level that we could select. It would be ideal for the rotational parameter to have some connection to intrinsic stellar properties. For instance, rotational velocity normalizes the rotational period to the stellar radius, providing a distinction between two main sequence stars of different masses that may have similar rotation periods. Optimally, the rotational variable in question would also provide a direct link to either observable magnetic activity or a theoretical description of stellar magnetism.

Accordingly, we advocate the use of the Rossby number (hereafter Ro). Ro is defined as the ratio of the rotational period of the star to its convective overturn time, Ro~=~$P_{\textrm{rot}}/\tau_{\textrm{conv}}$, and measures the strength of the Coriolis force acting on the vertical motion of convection cells. The dimensionless quantity Ro appears directly in standard mean-field dynamo theory \citep[$\alpha$-$\omega$ dynamo;][]{Parker1979}\footnote{For fully convective stars, an $\alpha$-$\omega$ dynamo cannot operate due to the lack of a tachocline. Instead, it is thought that an $\alpha^2$ dynamo can efficiently generate a magnetic field. Since it is the $\alpha$ mechanism that is related to the strength of the Coriolis force, the Rossby number should be just as applicable to fully convective stars \citep{Chabrier2006,Browning2008}.} and is intimately related to the ratio of the stellar x-ray luminosity to bolometric luminosity \citep{wright2011}. The latter quantity has been shown to be a strong indicator of a stellar corona heated to over $10^6$~K by magnetic activity \citep{Vaiana1981,Pallavicini1981,Noyes1984}. 

One cut in Ro was performed at Ro~=~0.1, illustrtated in Figure~\ref{fig:rossby}. The selection of Ro~=~0.1 approximately corresponds to Ro$_{\textrm{sat}}$, or the value of Ro associated with an observed saturation of the stellar dynamo apparent in the ratio of the stellar x-ray luminosity to the bolometric luminosity \citep[i.e., coronal saturation;][]{wright2011}. Intuitively, this suggests that all points with Ro values less than 0.1 are, presumably, sufficiently active so as to display inflated radii. Given our current understanding of coronal saturation, it is difficult to ascertain how strong of correlation is expected to exist. With that said, we assume that stars with very low Ro values should show at least a marginal degree of inflation compared to stars with higher Ro values, thereby indicating we should observe at least some evidence of a correlation.

There appears to be no emergent correlation between Ro and radius deviations, at least for the present sample of data. We find no significant difference in the distribution of data points  observed as a function of Ro. However, inspection of Figure~\ref{fig:rossby} highlights the need for more data in order to draw a definitive conclusion. Selection of Ro~=~0.1 has the unfortunate effect of creating a bin with a small sample population, potentially affecting the statistics. Field stars in wide binary systems are a good starting point for studies interested in populating the high Ro region of Figure~\ref{fig:rossby}. Caution must further be taken as secondary stars in the longer period ($>$~15d) systems in our sample do not have independently measured rotation periods, meaning they could potentially have different Ro values than are presented here. Visual inspection of Figure~\ref{fig:rossby} leads us to the same conclusion, if we ignore the two most discrepant points. Since we were comparing MAE values, the outliers do not significantly affect the results of the statistical analysis.

Arguably, the MAE is not an effective measure of the degree of \emph{inflation} of each sample as it treats deflated radii the same as inflated radii. Instead, the actual direct mean may provide a more compelling statistical measure. Thus, we ran our statistical analysis on the direct mean error. Overall, the typical degree of inflation among the radii of low-mass stars was found to be about 1.6\%. No statistically significant correlations with either $P_{\textrm{orb}}$ or Ro were uncovered. The most significant result was found for the period cut at 1.5d, where the K-S test indicated a significant difference. However, the MC method produced a difference in the mean error of the two populations at 2.2$\sigma$, below our significance threshold of 3.0$\sigma$. All other bin divisions yielded results significant at only about the 1$\sigma$ level.

Curiously, if we consider only data points with measured x-ray fluxes (see below), there is evidence that systems with lower Ro values may have larger radius discrepancies. However, we are then prompted to explain why the trend does not continue to higher Ro values, a question we are currently not equipped to answer. There is still some ambiguity with the presence of the points near Ro = 0.01, which appear to contradict the presence of any definite correlation. Deeper x-ray observations of x-ray faint low-mass DEBs will clarify this ambiguity.

\subsubsection{X-ray Activity}
\label{sec:xray}
An interesting extension of our discussion in the previous subsection, is to compare our derived radius deviations with the observed x-ray to bolometric luminosity ratio (hereafter $R_{x} = L_{x}/L_{bol}$). Since no correlation was noted with Ro, we expect that no correlation will be observed with $R_{x}$, as it has been shown to be intimately connected with Ro \citep{wright2011}. 

\citet{Lopezm2007} previously performed such a comparison and found a clear correlation between $R_{x}$ and radius deviations. Her comparison was performed under different modeling assumptions, which may have influenced the results. Specifically, she compared all observations to a 1~Gyr, solar metallicity isochrone from \citet{BCAH98}. Without variation in age and metallicity, the observed radius discrepancies may have been over estimated (or under estimated in the case of YY Gem). 

A subsample of sixteen DEBs from this study have identified x-ray counterparts in the \emph{ROSAT} All-Sky Survey Bright and Faint Source Catalogues \citep{Voges1999,Voges2000}. Our analysis follows that of \citet{Lopezm2007} whose previous analysis contained a fraction of our current x-ray detected sample. X-ray count rates were converted to x-ray fluxes according to the formula derived by \citet{Schmitt1995},
\begin{equation}
F_x = \left( 5.30 \textrm{HR} + 8.31\right)\times 10^{-12} X_{\textrm{cr}}
\end{equation}
where HR is the x-ray hardness ratio\footnote{There are typically two hardness ratios listed in the \emph{ROSAT} catalog, HR1 and HR2. The \citet{Schmitt1995} formula requires the use of HR1.}, $X_{\textrm{cr}}$ is the x-ray count rate, and $F_x$ is the x-ray flux. Luminosities in the x-ray spectrum were computed with distances determined from either \emph{Hipparcos} parallaxes \citep[preferred when available;][]{hip07} or near-infrared photometry. 

Photometric distances were estimated using the luminosity calculated from the Stefan-Boltzmann relation assuming the observationally determined radius and effective temperature. Absolute magnitudes were then derived using the PHOENIX AMES-COND model atmospheres, adopting the best fit isochrone metallicities. In an effort to reduce errors introduced by the theoretical atmospheres, distances were computed using the average distance modulus derived from \emph{2MASS} J and K band photometry \citep{2mass}.

We derived $R_x$ values for all sixteen stars in our x-ray sample to ensure internal consistency. Slight discrepancies between values presented here and those of \citet{Lopezm2007} are due entirely to differences in the adopted distances. Attributing the x-ray flux contribution from each star in an DEB system is a difficult task. As such, L\'{o}pez-Morales performed her analysis using three reported empirical scaling relations \citep{Pallavicini1981,Fleming1989}:
\begin{itemize}
\item Case 1 -- each component contributes equal weight.
\item Case 2 -- proportional to the respective rotational velocity, $v\sin i$, of each component.
\item Case 3 -- proportional to the square of the rotational velocity, $(v\sin i)^2$, of each component. 
\end{itemize}

We present results from all three cases in Figure~\ref{fig:xray}. Immediately we notice that the size of the stellar radius deviations appears to correlate with $R_x$. A linear least-square regression performed on each data set (light-blue dashed line in Figure~\ref{fig:xray}) suggests the slope for each case (1 -- 3) is 26, 24, and 22 ($\pm$ 17), respectively, all with reduced-$\chi^2$ values of $\sim$8. Pearson $r$ coefficients were 0.72, 0.69, and 0.63, respectively. The statistics suggest that the likelihood of uncorrelated sets of data producing these particular correlations are 0.2\%, 0.3\%, and 0.9\%, for case 1, 2, and 3, respectively. We can therefore rule out the null hypothesis with greater than 99\% confidence.

Visually, however, we notice that the correlation is largely driven by the presence of YY~Gem and UV~Psc~B, located in the upper-right region of each panel in Figure~\ref{fig:xray}. If we were to remove those three points (YY~Gem appears as a single point), the correlation vanishes and we only observe an offset from the zero point (Indigo dash-dotted line in Figure~\ref{fig:xray}). Furthermore, the distance to YY~Gem is highly uncertain. Assuming it is associated with Castor AB provides a distance of about 15~pc (see Appendix~\ref{sec:appendix}), but our photometric analysis, as described above, places YY~Gem at a distance of approximately 11~pc, 4~pc closer to the Sun than the Castor AB system. Instead of selecting a single distance estimate, we averaged the two estimates and adopted $d=13\pm 2$~pc, which also happens to be the distance assumed by \citet{Delfosse2003}.

The fact that our statistical correlation critically hinges upon three points, two of which are strongly distance dependent, is a cause for concern and implies that the statistics should be interpreted with care. If we believe the strong statistical correlation, then we are presented with a situation where the rotational data and the x-ray data disagree. This may be due to the physics underlying the stellar dynamo or those underlying x-ray saturation. However, there are two further interpretations that are contingent upon the role systematics play. 

First, we have that the correlation is entirely real and the presence of non-inflated stars with $R_x$~$\sim$~0.0007 are outliers in the relation. Second, systematics play a large role, as proposed in Section~\ref{sec:result}. Here, the truly deviant stars exhibit very strong x-ray emission ($R_x > 0.001$), while non-inflated stars show lower, varying levels of x-ray emission. For this view to hold, the relation between the level of radius inflation and magnetic field strength can not be linear. Strong magnetic fields would induce significant radius inflation while moderate and weak fields would produce little or no inflation.

Accepting, on the other hand, that the statistical correlation is spurious, we are left with a picture that is coherent with our rotation analysis. Namely, magnetic activity may not be the leading cause for all of the observed inflated radii. Here, systematics may still play a role in producing stars that appear inflated, but that are consistent with the models, leaving a few discrepant stars. Naturally inefficient convection, potentially dependent on particular stellar properties, may be operating. However, magnetic activity cannot be fully ruled out, as we do not yet have a fully self-consistent description of the interaction of magnetic fields with the stellar interior and atmosphere. It may be that magnetic fields acting within YY~Gem and UV~Psc have a more noticeable affect on stellar structure, as higher mass stars are affected more by changes to the properties of convection (see Appendix~\ref{sec:vmixl}).

We tend to favor a hybrid interpretation. Here, most of the observed ``inflation'' is an artifact of unaccounted-for systematics, but significantly discrepant stars (YY~Gem, UV~Psc) are associated with very strong magnetic activity. We believe that CM~Dra probably fits into the latter category due to a push in the literature towards subsolar metallicity. Effects of a large-scale magnetic field are presumably mass dependent, with higher mass stars showing a greater propensity to become inflated owing to their sensitivity to changes in convective properties. Whether there is a characteristic magnetic field strength that induces substantial radius deviations is unclear. Dynamo saturation and the saturation of magnetic activity, as evidenced by the flattening of the Ro-$R_x$ relation in \citet{wright2011}, is not yet fully understood, but may provide further insight into the apparent disagreement between our rotation and x-ray analyses.

Finally, clarity will be obtained with a better distance measurement to YY~Gem, either from ground-based parallax programs or with eventual results from \emph{Gaia}. Of all the points in Figure~\ref{fig:xray}, the existence of a positive correlation is most dependent upon the distance to YY~Gem. An accurate distance, as well as a reanalysis of its association with the Castor system, has the ability to not only relieve the ambiguity present in the x-ray data, but also to provide insight into the necessary constraints for the system (i.e., is YY~Gem really about 400 Myr old?). As further low-mass DEB systems are discovered, x-ray observations are strongly encouraged in order to develop a coherent picture of how radius deviations correlate with magnetic activity.

\section{Summary}
\label{sec:summary}
This study focused on reevaluating the current state of agreement between the theoretical and observational low-mass, main sequence MR relationship. The DEB systems used in the analysis were required to have quoted random uncertainties in the mass and radius below 3\% in order to provide an effective test of stellar evolution models. A large grid of DSEP models were computed with variation in age and metallicity characteristic of the local galactic neighborhood. Best fit isochrones were derived by allowing the age and metallicity to be optimized while maintaining the constraint that the system be coeval with a single composition. DEBs with reliably determined ages or metallicities were compared with a restricted set of isochrones in the range allowed by the observational priors.

Overall, we find 92\% of stars in our sample are less than 4\% discrepant with the models, largely representing a factor of two improvement over the canonical 5 -- 15\% deviations. Our results suggest that low-mass stars with radii that deviate significantly from model predictions are exceptions to general agreement. Discrepancies may also be the result of unaccounted for systematic uncertainties (i.e., starspots) that may as large as 4\%. With uncertainties as large as the typical radius deviations, we find it difficult to draw the firm conclusion suggesting that models are in broad disagreement with observations. Instead, we are left with a situation where the observational uncertainties may be too large to provide an adequate test of stellar models. The combination of random and systematic uncertainties for the sample of low-mass DEBs must be constrained and minimized below the 2\% level before accurate model comparisons may be made.

Radius correlations with orbital (rotational) period, Ro, and $R_{x}$ were also considered. No distinct trends were identified with either orbital period or Ro. However, we find evidence for a strong correlation between radius deviations and $R_{x}$ (previously noted by \citet{Lopezm2007}) in contradiction with our Ro analysis. The trend is not as tight as that derived by \citet{Lopezm2007}, owing to the age and metallicity variations allowed by our analysis. This correlation is also largely contingent upon the veracity of the distance estimate to YY~Gem. Accurately determining the distance to YY~Gem and evaluating its association with the Castor~AB quadruple would alleviate much of the uncertainty.   

Finally, we must not leave the theoreticians out of the spotlight. The degree to which a magnetic field can alter the interior structure of low-mass stars is still only partially known and further investigations are required. Development of models with self-consistent magnetic field perturbations will begin to shed light on this unknown. Comparisons between predicted magnetic field strengths from self-consistent models and observational data (either direct or indirect) will provide a measure of the validity of the ability of magnetic fields to inflate stellar radii. Whatever the final solution may be to this long-standing problem, it is now apparent that the level of inflation required by theory is not a ubiquitous 5~--~15\%, but only so in extreme cases.

\acknowledgements
We are grateful to G. Torres and J. Orosz for insightful discussions and access to preliminary data results. The authors would also like to extend gratitude to A. Irwin for his work on the open-source project FreeEOS. GAF wishes to thank the William H. Neukom 1964 Institute for Computational Science for their generous support. BC and GAF would both also like to acknowledge the support of the National Science Foundation (NSF) grant AST-0908345. This research has made use of NASA's Astrophysics Data System, the SIMBAD database, operated at CDS, Strasbourg, France, and the \emph{ROSAT} data archive tools hosted by the High Energy Astrophysics Science Archive Research Center (HEASARC) at NASA's Goddard Space Flight Center.

\appendix

\section{Variable Mixing-Length Models}
\label{sec:vmixl}

\subsection{Isochrone Grid}

A second set of isochrones was generated in order to address the idea that low-mass stars may possess inflated radii due to inefficient convective energy transport. It has been posited that convection within low-mass stars may either be inherently inefficient or that other physical processes (i.e., magnetic fields) act to reduce the ability of convection to effectively transport energy. We do not attempt to prescribe a physical mechanism associated with this conjecture. Instead, we attempt to parametrize convection in such a way so as to reduce the efficiency of convection in the lower mass regimes while still maintaining our solar calibration, necessary to properly model the Sun.

The second grid of isochrones was computed with a mass-dependent mixing-length, henceforth referred to as ``variable $\alpha_{MLT}$'' models. Convection for these models was parametrized with a smooth quadratic function of the form
\begin{equation}
\alpha\left(\frac{M}{M_{\odot}}\right) = a \left(\frac{M}{M_{\odot}}\right)^2 + b.
\end{equation}
Selection of a quadratic was arbitrary and carries no physical justification, except to produce low-mass stars with relatively inefficient convection compared to those in the standard model case.

Coefficients were determined by matching the convective mixing-length to predetermined values at two different masses. Since the overall structure of very low-mass stars is rather insensitive to the precise value of the mixing-length, we anchored $\alpha_{MLT} = 1.00$ at $M = 0.1M_{\odot}$. The other end of the mass spectrum is constrained by our need to satisfy our solar calibration. Thus, at $M = 1.0M_{\odot}$, the convective mixing-length was fixed to $\alpha_{MLT} = 1.94$. Subsequently,
\begin{equation}
\alpha\left(\frac{M}{M_{\odot}}\right) = 0.949 \left(\frac{M}{M_{\odot}}\right)^2 + 0.991.
\label{eq:alpha_mod}
\end{equation}
A relative comparison between models computed with a solar calibrated mixing-length and those generated with a parametrized mixing-length is presented in Figure~\ref{fig:mixl}~A.

Our isochrone grid for these models covered a fraction of the parameter space compared to the standard model set. There was a total of 12 isochrones for the variable $\alpha_{MLT}$ models: [Fe/H] = \{-0.5, 0.0, +0.2\} dex with ages = \{1.0, 3.0, 5.0, 8.0\} Gyr. Below the fully convective boundary, the models are rather insensitive to the mixing-length, as stated above. For higher masses, the reduced mixing-length increases the model radii by up to 3\%. As designed, a Sun-like star is unaffected by this mixing-length prescription. We also see that above $1M_{\odot}$ the model radii begin to decrease due to an increased mixing-length as a result of our parametrization.
\begin{deluxetable*}{l c r r r r c c r r r r}[hb]
\tablecaption{Best fit isochrone with a mass-dependent convective mixing-length.}
\tablecolumns{12}
\tablehead{ 
 \colhead{} &  \multicolumn{5}{c}{Quoted} & & \multicolumn{5}{c}{Fixed 3\%} \\
 \cline{2-6}\cline{8-12} \\
 \colhead{Star Name} & \colhead{Age} & \colhead{[Fe/H]} & \colhead{$\delta R/R_{\textrm{obs}}$} & \colhead{\# $\sigma_{R}$} & \colhead{Fit} & & 
 \colhead{Age} & \colhead{[Fe/H]} & \colhead{$\delta R/R_{\textrm{obs}}$} & \colhead{\# $\sigma_{R}$} & \colhead{Fit}
}
\startdata
UV Psc A          & 8.0 &  0.00 & -0.0072 &  -0.349 &  No & & 8.0 &  0.00 & -0.0072 &  -0.241 &   No \\
UV Psc B          &     &       &  0.0866 &   4.019 &     & &     &       &  0.0866 &   2.888 &      \\
IM Vir A          & 8.0 & +0.20 & -0.0032 &  -0.209 &  No & & 8.0 & +0.20 & -0.0032 &  -0.105 &  Yes \\
IM Vir B          &     &       &  0.0238 &   1.245 &     & &     &       &  0.0238 &   0.792 &      \\
KID 6131659 A     & 3.0 & -0.50 & -0.0113 &  -5.876 &  No & & 3.0 & -0.50 &  0.0113 &  -0.378 &  Yes \\
KID 6131659 B     &     &       & -0.0190 &  -9.352 &     & &     &       & -0.0190 &  -0.634 &      \\
RX J0239.1-1028 A & 8.0 & -0.50 & -0.0075 &  -1.394 &  No & & 8.0 & -0.50 & -0.0075 &  -0.251 &  Yes \\
RX J0239.1-1028 B &     &       &  0.0039 &   1.356 &     & &     &       &  0.0039 &   0.129 &      \\
Kepler-16 A       & 1.0 &  0.00 & -0.0085 &  -4.227 &  No & & 1.0 &  0.00 & -0.0085 &  -0.282 &  Yes \\
Kepler-16 B       &     &       &  0.0244 &  11.061 &     & &     &       &  0.0245 &   0.815 &      \\
GU Boo A          & 8.0 & -0.50 &  0.0085 &   0.333 & Yes & & 8.0 & -0.50 &  0.0085 &   0.283 &  Yes \\
GU Boo B          &     &       &  0.0190 &   0.740 &     & &     &       &  0.0190 &   0.632 &      \\
YY Gem A          & 0.4 &  0.00 &  0.0757 &   8.226 &  No & & 0.4 &  0.00 &  0.0757 &   2.523 &   No \\
YY Gem B          &     &       &  0.0757 &   8.226 &     & &     &       &  0.0757 &   2.523 &      \\
MG1-506664 A      & 1.0 &  0.00 & -0.0017 &  -0.964 &  No & & 1.0 &  0.00 & -0.0017 &  -0.057 &  Yes \\
MG1-506664 B      &     &       & -0.0099 &  -5.065 &     & &     &       & -0.0099 &  -0.329 &      \\
MG1-116309 A      & 8.0 & +0.20 & -0.0126 &  -1.743 &  No & & 8.0 & +0.20 & -0.0126 &  -0.421 &  Yes \\
MG1-116309 B      &     &       &  0.0115 &   1.534 &     & &     &       &  0.0115 &   0.384 &      \\
MG1-1819499 A     & 5.0 & -0.50 &  0.0240 &   6.839 &  No & & 5.0 &  0.00 &  0.0413 &   1.376 &   No \\ 
MG1-1819499 B     &     &       & -0.0575 &  -9.587 &     & &     &       & -0.0396 &  -1.320 &      \\
NSVS 01031772 A   & 8.0 & +0.20 & -0.0121 &  -2.276 &  No & & 8.0 & +0.20 & -0.0121 &  -0.404 &   No \\
NSVS 01031772 B   &     &       &  0.0352 &   5.965 &     & &     &       &  0.0352 &   1.172 &      \\
MG1-78457 A       & 3.0 &  0.00 & -0.0023 &  -0.143 & Yes & & 3.0 &  0.00 & -0.0023 &  -0.075 &  Yes \\
MG1-78457 B       &     &       &  0.0005 &   0.025 &     & &     &       &  0.0005 &   0.016 &      \\
MG1-646680 A      & 1.0 &  0.00 & -0.0319 &  -2.428 &  No & & 1.0 &  0.00 & -0.0319 &  -1.063 &   No \\
MG1-646680 B      &     &       &  0.0159 &   1.129 &     & &     &       &  0.0159 &   0.529 &      \\
MG1-2056316 A     & 1.0 & +0.20 & -0.0083 &  -1.829 &  No & & 1.0 & +0.20 & -0.0083 &  -0.277 &  Yes \\
MG1-2056316 B     &     &       &  0.0112 &   2.095 &     & &     &       &  0.0112 &   0.373 &      \\
CU Cnc A          & 8.0 & +0.20 &  0.0127 &   0.998 & Yes & & 8.0 & +0.20 &  0.0127 &   0.423 &  Yes \\
CU Cnc B          &     &       & -0.0028 &  -0.116 &     & &     &       & -0.0028 &  -0.092 &      \\
LSPM J1112+7626 A & 8.0 & +0.20 & -0.0038 &  -0.278 &  No & & 8.0 & +0.20 & -0.0038 &  -0.125 &  Yes \\
LSPM J1112+7626 B &     &       &  0.0255 &   1.652 &     & &     &       &  0.0255 &   0.850 &      \\
KOI-126 B         & 3.0 & +0.20 &  0.0020 &   0.365 & Yes & & 5.0 &  0.00 &  0.0010 &   0.033 &  Yes \\
KOI-126 C         &     &       & -0.0017 &  -0.307 &     & &     &       & -0.0014 &  -0.047 &      \\
CM Dra A          & 5.0 &  0.00 &  0.0282 &   3.757 &  No & & 5.0 &  0.00 &  0.0282 &   0.939 &   No \\
CM Dra B          &     &       &  0.0333 &   4.441 &     & &     &       &  0.0333 &   1.111 &     
\enddata
\label{tab:nsmixl}
\end{deluxetable*}

\subsection{Results}

\begin{figure*}
\plotone{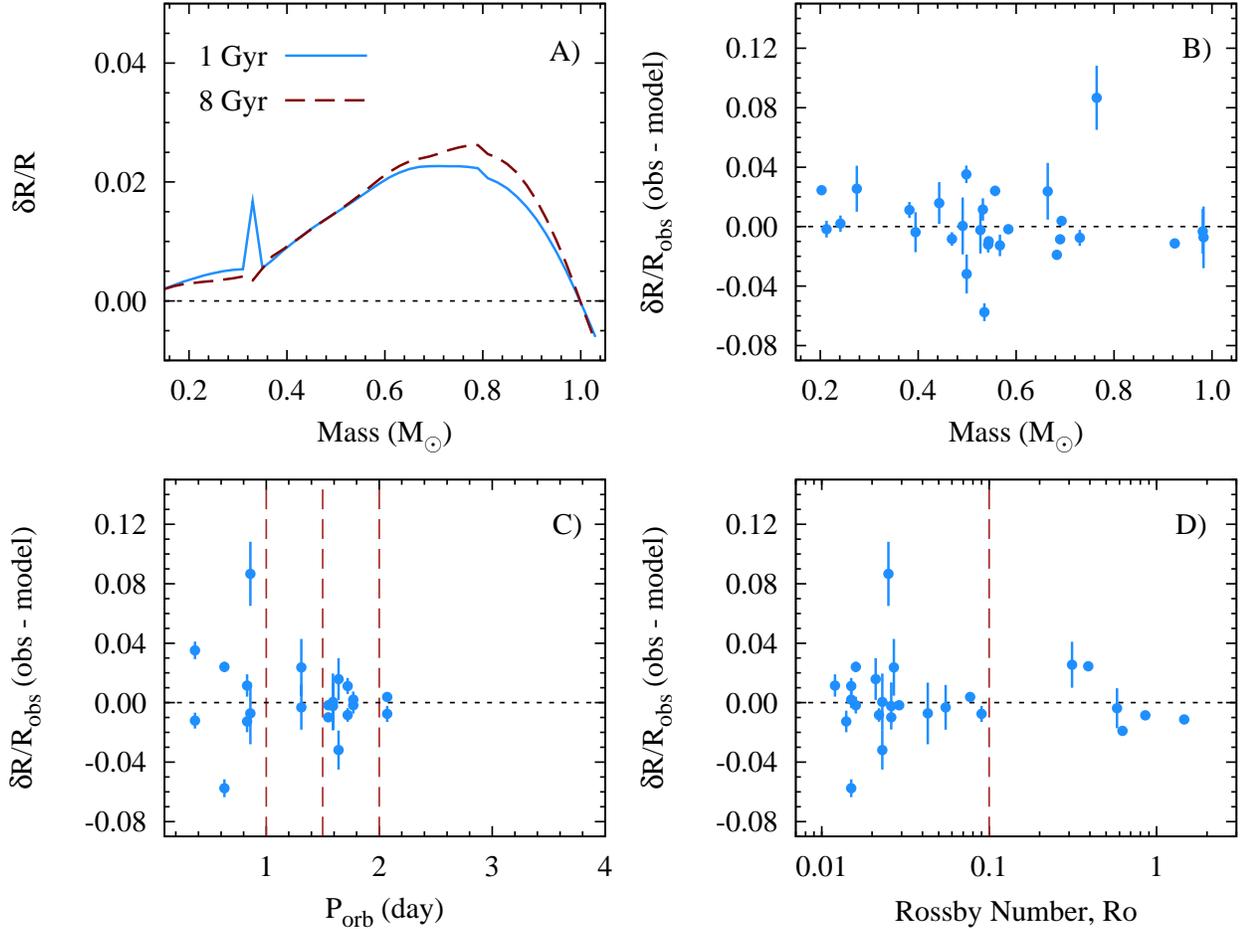}
\caption{\emph{Top-left:} Radius variations induced by our prescription of a mass-dependent $\alpha_{MLT}$ (Eq.~\ref{eq:alpha_mod}). Shown are 1 Gyr (blue -- solid) and 8 Gyr (maroon -- dashed) isochrones computed with a solar heavy element composition (GS98). Positive values indicate that the modified $\alpha_{MLT}$ models are inflated compared to the solar calibrated models. The ``blip'' observed near the fully convective boundary is most likely due to a $^3$He instability described by \citet{vSP2012}. \emph{Top-right:} Same as Figure~\ref{fig:drad}~A excepted for the modified $\alpha_{MLT}$ models. \emph{Bottom-left:} Same as Figure~\ref{fig:porb}~B, although we only present the short period systems. The distribution of long period systems is similar to Figure~\ref{fig:porb}~A. \emph{Bottom-right:} Identical to Figure~\ref{fig:rossby}. }
\label{fig:mixl}
\end{figure*}

Non-standard, variable mixing-length models lead to only slightly better results over the standard models when considering the quoted random uncertainties (see Table~\ref{tab:nsmixl}). We again find that radius deviations are largely reduced to below 4\%, seen clearly in Figure~\ref{fig:mixl}~B. The residuals do appear to be more tightly clustered around the zero point, although only slightly, suggesting an overall better agreement between models and observations. This is reinforced by a slightly lower MAE of 2.1\% across the entire sample.

There was evidence in Figure~\ref{fig:mixl}~B of a possible tendency for the best fit theoretical isochrone to over predict the observed radius at higher masses and under predict the radius for lower mass stars. A least squares regression on the data finds a linear slope of $-0.031~\pm~0.013$, significant at the $2.5\sigma$ level (not shown). Specifically, half of the DEB systems that were not formally fit are characterized by an isochrone that over-predicts the higher mass primary but under-predicts the lower mass secondary. This is only an artifact of our functional parametrization of $\alpha_{MLT}$, suggesting convection was too heavily suppressed at higher masses and under-suppressed in the low-mass regime.

We subjected the resulting data to the same statistical analyses described in Section~\ref{sec:result1}. We again compared the radius deviations to the rotational period and Rossby number (Figure~\ref{fig:mixl}~C and D). The strongest hint of a correlation was found when applying a cut at a period of 1.5d, although it was only significant at the $2.7\sigma$ level, below our $3\sigma$ significance threshold. Performing the period cut at 1.0d and 2.0d yielded results that were even less significant (2.4$\sigma$ and 1.9$\sigma$, respectively), providing the same qualitative result as standard model scenario. Similarly, there is no strong correlation observed between the radius deviations and the Rossby number.

Again, artificially fixing the uncertainties at 3\% tends to add ambiguity as to whether there are any real discrepancies for a majority of the systems. As was suggested throughout our study of the standard models, systematic uncertainties must be minimized to below 2\% before an accurate comparison between models and observations may occur. 

\section{Age and Metallicity Priors}
\label{sec:appendix}
Mentioned in Section~\ref{sec:prior}, it was determined that four of the DEBs in our sample had age and metallicity priors reliably determined. These four were a subset of a total of eight DEB systems that had either an estimated age or metallicity in the literature. Below, we provide our analysis and reasoning for our acceptance or rejection of each quoted prior.

\subsection{UV Psc}
The age of the UV Psc system is quoted by the TAG10 review to be 7.9 Gyr. Popper's original paper describing the characteristics of the UV Psc system does not provide any evidence to support an age estimate \citep{Popper1997}. An age of approximately 8 Gyr is typically assigned to the system due to the fact that stellar evolution models predict the physical properties of the primary star, UV Psc A, at that age. However, there has yet to be any set of models that can place both components on a consistent, coeval isochrone. Since the estimated age of UV Psc A appears to be derived from stellar evolution models, we allowed our models to independently determine the most acceptable age. 

\subsection{IM Vir}
\citet{Morales2009b} attempt to determine the metallicty of IM Vir by applying various photometric metallicity relations. The result of their efforts was that they found all of the various empirical methods quote different values with large uncertainties. Values for [Fe/H] vary from the metal-poor end with [Fe/H] = -0.8 up to a super-solar value of [Fe/H] = +0.15. This range also fits nicely within the set of model metallicity values selected for this study. Since the cited metallicity range would not provide any additional constraint on our analysis, we rejected the metallicity prior.

\subsection{Kepler-16}
Kepler-16, the first confirmed binary system with a circumbinary planet, was provided with a metallicity estimate in its discovery paper \citep{Doyle2011}. A metallicity of [Fe/H] = -0.3$\pm$0.2 was determined spectroscopically. The authors indicated the spectroscopic analysis was performed on the K-dwarf primary and that the general method was similar to that applied to KOI-126 using Spectroscopy Made Easy \citep[SME;][]{sme, Carter2011}. Reliability is lent to the method, in general, due to its success at deriving the metallicity of KOI-126 A. We therefore accepted the quoted metallicity prior.

\subsection{GU Boo}
The age estimate provided by \citet{Lopezm2005} in their characterization of GU Boo was primarily based on kinematics. Specifically, they conclude that GU Boo is an isolated system and the vertical component of its motion provides a hint that it has undergone perturbations due to disk heating processes. Assuming that the system has been subjected to disk heating, one can only infer that the system has an age greater then $10^{8}$ yr, the typical timescale for dynamical perturbations associated with an objects orbit around the galactic center \citep{Soderblom2010}. Unfortunately, no further constraints were able to be placed on the age of the system. However, it is not possible to rule out the scenario that GU Boo was dynamically ejected from its stellar nursery. It would then appear that no reliable age estimation exists, prompting us to reject the age prior for this system.

\subsection{YY Gem}
YY Gem is thought to be physically associated with the Castor AB ($\alpha$ Gem) quadruple system. All three systems were found to be gravitationally bound based on a statistical analysis performed using a three-body interaction code \citep{aoc89}. While the \citet{aoc89} analysis was performed with pre-\emph{Hipparcos} proper motions as their initial conditions, it is unlikely that the results will be effected at the level necessary to unbind the systems.

Castor A and B are themselves both binaries. The primary in both systems is an A star and both are thought to have an M dwarf companion. Therefore, physical properties derived from spectroscopic and photometric observations of the two A stars will be essentially unaffected by the presence of their companions. Placing the two primaries on an M$_{\textrm{V}}$-Log($T_{\textrm{eff}}$) H-R diagram, we used DSEP to derive an age of about 400 Myr. This age is consistent with the average age of 370 Myr derived by \citet{Torres2002}, who modeled the A-star primaries using multiple stellar evolution codes. 

The A stars also lend themselves well to a spectroscopic determination of the metallicity. Unfortunately, there appears to be only one result reported. \citet{Smith1974} estimates an average metallicity for the two A stars to be about [Fe/H] = +0.7 relative to Vega. As described in \citet{Torres2002}, this implies a rather uncertain metallicity relative to the Sun of [Fe/H] = +$0.1\pm0.2$. Despite this, we adopt this metallicity constraint due to its rather large uncertainty, which should presumably encompass the true value.

\subsection{CU Cnc}
When CU Cnc was originally investigated by \citet{Ribas2003}, it was found to have a space motion very similar to that of the Castor sextuple system ($\alpha$ Gem, YY Gem) and, subsequently, the proposed ``Castor moving group.'' For this reason, CU Cnc was deemed to be associated with the Castor moving group implying an age and metallicity similar to the Castor sextuple. However, determining membership of a moving group is complicated and has often lead to ambiguous results concerning the coeval nature of the group. It is not uncommon for members of the same kinematic moving group to have different metallicities, implying that members of a defined moving group may not have been born in the same galactic environment and, as such, are not necessarily coeval. The lack of an age estimation beyond its kinematic similarity to the Castor system led us to reject the age prior of CU Cnc.

\subsection{KOI-126}
KOI-126 is a hierarchical triple eclipsing binary recently whose discovery was recently announced \citep{Carter2011}. The quoted metallicity prior of [Fe/H] = +$0.15\pm0.08$ was determined using SME, as was mentioned above in the discussion of Kepler 16. Assuming this metallicity allowed for a relatively precise age constraint ($\sim$4$\pm$1 Gyr) to be placed on the primary, a 1.35 $M_{\odot}$ subgiant. Combining both the age and metallicity information led to two low-mass companions to also be fit by stellar models \citep{Feiden2011}. Thus, there appears to be little question about the validity of the age and metallicity estimations, leading us to adopt the given priors for our study.

\subsection{CM Dra}
Finally, we consider the very well studied CM Dra system. Spectroscopic observations of the system have produced varying results for the metallicity of the system, but all appear to be consistent with -1 $\le$ [Fe/H] $\le$ 0 \citep{Viti1997,Viti2002,Morales2009a,Kuznetsov2012}. Due to the difficulties in modeling, and thus fitting, the entire SED of an M dwarf, we determined there was no particular reason to strongly favor one metallicity result over another. 

An age was determined for the system through the use of white dwarf (WD) cooling tracks combined with stellar evolution models of the approximate WD progenitor \citep{Morales2009a}. The WD cooling age was found to be 2.38$\pm$0.37 Gyr. \citet{Morales2009a} predicted a mass for the progenitor star of $M=2.1\pm0.4\, M_{\odot}$ and derived a total (stellar model + WD cooling) age of 4.1$\pm$0.8 Gyr. For consistency, we calculated the lifetime of the progenitor star using DSEP with the full available suite of physics. Accounting for the metallicity constraint defined above, we derived an age of 3.7$\pm$1.2 Gyr and, in doing so, found no reason to reject, or modify, the age prior.

\end{document}